\documentclass[a4paper,11pt]{article}
\pdfoutput=1

\usepackage{jheppub}
\usepackage{amsmath}
\usepackage{amssymb}
\usepackage[normalem]{ulem} 

\allowdisplaybreaks 
\def\Fig#1{fig.~{\ref{#1}}}
\DeclareRobustCommand{\Sec}[1]{sec.~\ref{#1}}
\DeclareRobustCommand{\Eq}[1]{eq.~(\ref{#1})}

\newcommand{\f}{\frac}
\newcommand{\df}{\mathrm{d}}

\newcommand{\nn}{\nonumber}
\newcommand{\nnu}{\nonumber\\}

\def\bnslash{\bar n\!\!\!\slash}
\preprint{MIT-CTP-5589 / JLAB-THY-23-3891}

\title{A Formalism for Extracting Track Functions from Jet Measurements}
\author[a]{Kyle Lee,}
\affiliation[a]{Center for Theoretical Physics, Massachusetts Institute of Technology, Cambridge, MA 02139, USA}

\author[b]{Ian Moult,}
\affiliation[b]{Department of Physics, Yale University, New Haven, CT 06511}

\author[c,d]{Felix Ringer,}
\affiliation[c]{Department of Physics, Old Dominion University, Norfolk, VA 23529, USA}
\affiliation[d]{Thomas Jefferson National Accelerator Facility, Newport News, VA 23606, USA}

\author[e,f]{Wouter J. Waalewijn}
\affiliation[e]{Institute for Theoretical Physics Amsterdam and Delta Institute for Theoretical Physics, University of Amsterdam, Science Park 904, 1098 XH Amsterdam, The Netherlands}
\affiliation[f]{Nikhef, Theory Group, Science Park 105, 1098 XG, Amsterdam, The Netherlands}

\emailAdd{kylel@mit.edu, ian.moult@yale.edu, fmringer@jlab.org, w.j.waalewijn@uva.nl}

\abstract{
The continued success of the jet substructure program will require widespread use of tracking information to enable increasingly precise measurements of a broader class of observables.
The recent reformulation of jet substructure in terms of energy correlators has simplified the incorporation of universal non-perturbative matrix elements, so called ``track functions", in jet substructure calculations.
These advances make it timely to understand how these universal non-perturbative functions can be extracted from hadron collider data, which is  complicated by the use jet algorithms.
In this paper we introduce a new class of jet functions, which we call (semi-inclusive) track jet functions, which describe measurements of the track energy fraction in identified jets.
These track jet functions can be matched onto the universal track functions, with perturbatively calculable matching coefficients that incorporate the jet algorithm dependence.
We perform this matching, and present phenomenological results for the charged energy fraction in jets at the LHC and EIC/HERA at collinear next-to-leading logarithmic accuracy.
We show that higher moments of the charged energy fraction directly exhibit non-linear Lorentzian renormalization group flows, allowing the study of these flows with collider data.
Our factorization theorem enables the extraction of universal track functions from jet measurements, opening the door to their use for a precision jet substructure program.
}

\begin{document}

\maketitle

\section{Introduction}

Jet substructure \cite{Larkoski:2017jix,Kogler:2018hem} plays a central role in many aspects of collider physics, from searches for new physics, to precision Standard Model measurements, to studies of QCD. To further extend the ability to decode energy flux within jets will require the study of higher-point correlations, as well as higher precision. Experimentally, this will require an increased use of tracking information, which has already proven crucial in cutting-edge jet substructure measurements at ATLAS \cite{STDM-2010-14,STDM-2014-17,STDM-2017-16,STDM-2017-33,STDM-2018-57}, CMS~\cite{CMS-SMP-20-010,CMS:2023ovl}, ALICE~\cite{ALargeIonColliderExperiment:2021mqf,ALICE:2019ykw,ALICE:2021njq,ALICE:2021aqk} and LHCb~\cite{LHCb:2017llq,LHCb:2019qoc}, as well as in many measurements of fragmentation \cite{ATLAS:2011myc,CMS:2014jjt,ALICE:2014dla,ATLAS:2017pgl,ALICE:2018ype,LHCb:2019qoc,ATLAS:2019dsv}. Since tracking information uses the charges of hadrons, theoretical calculations of jet substructure observables on tracks are not infrared and collinear safe \cite{Kinoshita:1962ur,Lee:1964is}, and hence cannot be performed purely in perturbation theory. This significantly complicates their theoretical description, and until recently has acted as a blockade to achieving precision theoretical calculations of jet substructure observables on tracks.

In ref.~\cite{Chang:2013rca,Chang:2013iba} a factorization for observables measured on tracks was introduced, allowing a separation of perturbative physics, whose description can be systematically improved by higher-order calculations, from non-perturbative physics, described by universal non-perturbative matrix elements. The non-perturbative inputs appearing in this factorization  were termed ``track functions" \cite{Chang:2013rca,Chang:2013iba,Chen:2022pdu,Chen:2022muj,Jaarsma:2022kdd,Li:2021zcf}. While originally formulated in the context of incorporating tracking information, this formalism applies much more generally for computing jet substructure observables  on subsets of final-state hadrons.

Due to a recent reformulation of jet substructure \cite{Dixon:2019uzg,Chen:2020vvp} in terms of energy correlators \cite{Basham:1979gh,Basham:1978zq,Basham:1978bw,Basham:1977iq,Hofman:2008ar} (for applications see refs.~\cite{Lee:2022ige,Komiske:2022enw,Holguin:2022epo,Liu:2022wop,Liu:2023aqb,Cao:2023rga,Devereaux:2023vjz,Andres:2022ovj,Andres:2023xwr,Craft:2022kdo,Andres:2023ymw}), the use of track functions in higher-order perturbative calculations has now become practical. This has reinvigorated their study, leading to a systematic understanding of their renormalization group evolution, and explicit calculations to next-to-leading order \cite{Chen:2022pdu,Chen:2022muj,Jaarsma:2022kdd,Li:2021zcf}. These calculations revealed an interesting non-linear renormalization group structure going beyond the standard linear Dokshitzer-Gribov-Lipatov-Altarelli-Parisi (DGLAP) \cite{Gribov:1972ri,Dokshitzer:1977sg,Altarelli:1977zs} paradigm of fragmentation functions (FFs) and parton distribution functions (PDFs). Efficient numerical implementations of the track function evolution were developed in refs.~\cite{Chen:2022pdu,Chen:2022muj}, and are available at \url{https://github.com/HaoChern14/Track-Evolution}, making them ready for phenomenological applications. Indeed, they have recently been applied to compute the small angle limit of the energy correlators on tracks at collinear next-to-leading logarithmic accuracy \cite{Yibei:new}.

While the renormalization group equations of the track functions can be systematically computed in perturbation theory, the values of the track functions at a particular scale are non-perturbative  parameters of QCD. Since track functions describe timelike dynamics near the lightcone, there is currently no known way to compute them from first principles. This is in contrast to PDFs, which describe spacelike dynamics, and for which there has been significant recent progress in calculations using Euclidean lattice formulations of QCD (see e.g.~refs.~\cite{Detmold:2001dv,Ji:2013dva,Ji:2014gla,Izubuchi:2018srq,Ji:2020ect}). Therefore, the missing link in the application of track functions to improving jet substructure measurements is an understanding of how they can be extracted from data. This is similar to the extraction of the universal FFs (see e.g.~\cite{deFlorian:2007aj,Bertone:2017tyb,Moffat:2021dji,Borsa:2022vvp}) and PDFs \cite{Pumplin:2002vw,Gao:2013xoa,Lai:2010vv,NNPDF:2014otw} from hadron collider data using the foundational factorization theorems of Collins-Soper-Sterman \cite{Collins:1989gx,Collins:1988ig,Collins:1985ue}. Despite their more complicated RG evolution, we actually think that the extraction of track functions from data will be easier than FFs, due to their simpler behavior at small momentum fraction. Due to the increasing importance of jet substructure in hadron collider physics, we hope that track functions can soon be found alongside PDFs and FFs as standard universal non-perturbative inputs for collider physics.

Track functions describe the \emph{total} fraction of energy carried by charged hadrons from the fragmentation of a quark or gluon. In $e^+e^-$ colliders they can be extracted by measuring the fraction of energy in charged hadrons. This was computed to NNLO in \cite{Chen:2022muj}, and was seen to exhibit excellent perturbative convergence. However, at hadron colliders, where we currently have the most precise data, the situation is more complicated due to the necessity of jet algorithms \cite{Ellis:1993tq,Cacciari:2005hq,Cacciari:2008gp,Cacciari:2011ma} to identify hard scattering events. In this case, the practical measurement that can be made is the energy fraction, $x_\text{trk}$, of charged hadrons inside a jet defined by a given jet algorithm and transverse momentum $p_T$. This is significantly more complicated both experimentally and theoretically. From the experimental perspective, this measurement is challenging since it requires a precise measurement of both the charged energy flux and total energy flux within identified jets. The measurement of the total energy flux is difficult, with a more limited jet energy scale resolution at the LHC~\cite{PERF-2011-05,PERF-2016-04,PERF-2015-05,JETM-2018-05}. This is in contrast to $e^+e^-$, where the hard scale $Q$ is fixed, and so only the energy of charged particles needs to be measured. From the theoretical perspective, this observable is also more complicated, since unlike the track functions themselves, these measurements are non-universal, exhibiting an explicit dependence on the jet algorithm.

In this paper, we derive a factorization theorem to relate measurements of the track energy fraction inside high-$p_T$ jets to universal track functions, enabling the extraction of universal track functions from hadron-collider data. Our factorization theorem involves a new type of jet function, which we term (semi-inclusive) track jet functions due to their analogy with their (semi-inclusive) fragmenting jet functions' counterparts~\cite{Kang:2020xyq,Kang:2016ehg}. These track jet functions can be matched onto the standard track functions, with perturbatively calculable coefficients that incorporate the details of the jet algorithm. We compute these objects for anti-$k_T$ \cite{Cacciari:2008gp} jets, and study their properties. We also show that they are a gen-
erating functional for (semi-inclusive) multi-hadron fragmenting jet functions, much like how track functions are a generating functional for multi-hadron fragmentation functions \cite{Chen:2022pdu,Chen:2022muj}. We then present phenomenological results both at the LHC, and at deep-inelastic scattering experiments (HERA/EIC), and show the importance of the track jet functions for properly extracting the universal track functions from jet-based measurements. Our results bridge recent theoretical advances in our understanding of track functions with experimental data, and provide the final missing piece for extracting universal track functions from jet measurements (apart from the measurement itself!).

There are two primary reasons why measurements of the track functions are interesting. The first is, as mentioned above, that there is currently no way to calculate the track function from first principles, and therefore it must be measured if it is to be used in precision jet substructure calculations. Its universality then ensures that it can be used in any experiment. This is similar to the program of extracting PDFs or FFs. This is an extremely valuable output of the measurement of the track function, and will certainly lead to new tests of QCD from new jet substructure measurements. However, due to the fact that there is currently no first principles way of computing the track function from the QCD Lagrangian, the measurement of the track function at a single scale does not provide a test of QCD. The second motivation for measuring the track functions is as a test of the dynamics of QCD, due to the fact that the renormalization group evolution of the track functions can be computed in perturbative QCD, and exhibits remarkable features of QCD that to our knowledge have not previously been directly measured.

A particular interesting aspect of the measurement of the $x_\text{trk}$ is that it provides not just a value of the mean $\langle x_\text{trk} \rangle$, but also of higher moments, $\langle (x_\text{trk}-\langle x_\text{trk} \rangle)^n \rangle$, which encode interesting fluctuations in the fragmentation process. While the mean value of fragmentation quantities, such as the multiplicity can be described to high orders \cite{Kotikov:2004er}, fluctuations are much less understood, and are one of the primary advances arising from the study of jet substructure. Higher moments of the energy fraction distribution in jets exhibit non-linear RG flows of the underlying track functions, allowing these RG flows to be directly studied in a clean experimental observable. It is precisely these higher moments, which capture fluctuations as opposed to mean values, that make jet substructure theoretically interesting, and is where the advances in our understanding using field theory, have occurred. The non-linearity of the RG is a feature common to a number of \emph{Lorentzian} RG equations, such as the BK \cite{Balitsky:1995ub,Kovchegov:1999yj} or B-JIMWLK \cite{JalilianMarian:1997jx,Iancu:2000hn} equations for forward scattering, the BMS equation for the resummation of non-global logarithms \cite{Dasgupta:2001sh,Banfi:2002hw}, the evolution equations for leading jets \cite{Scott:2019wlk,Neill:2021std}, and small-z fragmentation \cite{Neill:2020bwv,Neill:2020tzl}. However, a unique feature here is that we are able to directly measure the RG flows of these operators. This is, to our knowledge, unique, and provides interesting insight into Lorentzian RG phenomena.

An outline of this paper is as follows: In \Sec{sec:semi}, we introduce the track jet functions, and discuss some of their basic properties, including their operator product expansion onto the universal track function, their renormalization group evolution, and their relation to (semi-inclusive) multi-hadron fragmenting jet functions. In \Sec{sec:pert}, we study the perturbative structure of the track jet functions, computing them at one-loop, and predicting their two-loop structure using the renormalization group. We then present numerical results for the energy fraction on charged hadrons inside identified jets at the LHC and HERA/EIC in \Sec{sec:pheno}. We highlight the large numerical impact of the matching coefficients in properly extracting the universal track functions, as well as the interesting renormalization group flows that can be probed through measurements of the $x_\text{trk}$ distribution. We conclude in \Sec{sec:conc}.

\section{Track Jet Functions}\label{sec:semi}

In this section we define the track jet functions and study their basic properties, namely their operator product expansion onto universal track functions, their renormalization group evolution, and the structure of their moments. We also show that they are a generating functional for (semi-inclusive) multi-hadron fragmenting jet functions.

\subsection{Definition}\label{sec:def}

\begin{figure}
\begin{center}
\includegraphics[scale=0.35]{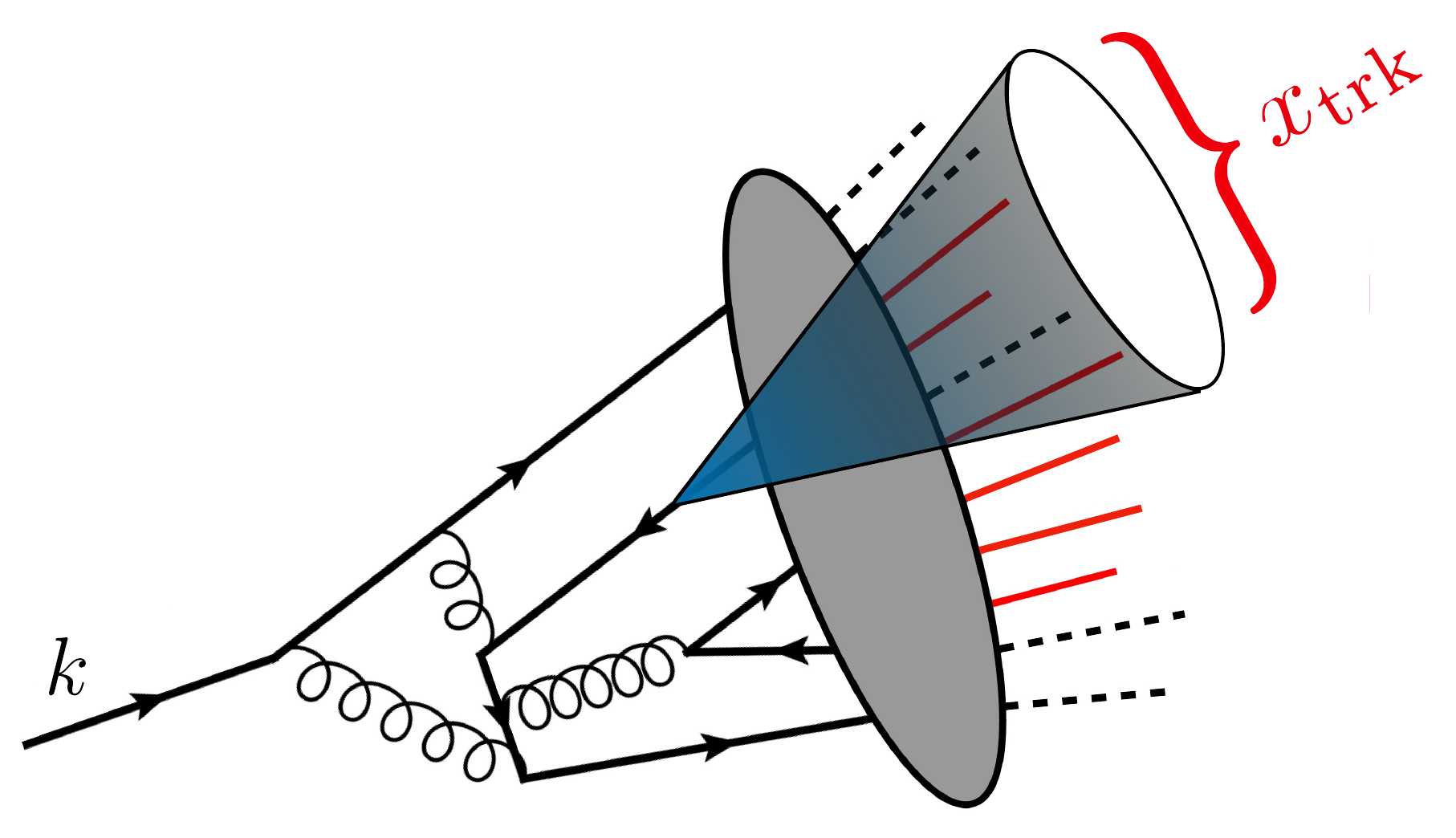} 
\end{center}
\caption{An illustration of a jet with charged momentum fraction $x_\text{trk}$, produced through the fragmentation of a quark with momentum $k$. Charged particles are indicated in red, and the grey blob denotes the hadronization process from partons to hadrons. This is described field-theoretically using a track jet function.}
\label{fig:ft}
\end{figure}

Our goal is to describe the momentum fraction of tracks with respect to the momentum of an identified anti-$k_T$ jet, which we will denote by $x_{\rm trk}$. We illustrate this in \Fig{fig:ft}. In particular, we want to express the cross section for this observable in terms of universal track functions, which describe the non-perturbative dynamics at the scale $\Lambda_{\text{QCD}}$, and a perturbative matching coefficient, which describes the jet formation at the scale $p_T R$, where $p_T$ is the transverse momentum of the jet and $R$ is its radius. This is a generalization of factorization theorems for identified hadrons in jets, first considered in  ref.~\cite{Procura:2009vm}, and significantly developed and extended to a variety of different contexts~\cite{Jain:2011xz,Procura:2011aq,Kaufmann:2015hma,Dai:2016hzf,Dai:2017dpc,Kang:2016ehg,Kang:2017frl,Kang:2020xyq}.

To obtain a factorization theorem for the $x_{\rm trk}$ observable in a hadron collider environment, we begin with the factorization theorem of Collins-Soper-Sterman \cite{Collins:1981ta,Bodwin:1984hc,Collins:1985ue,Collins:1988ig,Collins:1989gx,Nayak:2005rt,Collins:2011zzd} for inclusive hadron production
\begin{align}
\label{eq:fact}
\frac{\mathrm{d} \sigma}{\mathrm{d} p_{T}\,\mathrm{d} \eta\, \mathrm{d} x_{\rm trk}}= \mathcal{H}_{i}(p_{T}/z,\eta, \mu)\otimes_z  \mathcal{G}_{i\to \rm{trk}}(z,x_{\rm trk},p_T R,\mu)\,,
\end{align}
where the fragmentation function is replaced by the track jet function $\mathcal{G}$, which describes the inclusive production of jets from an initial parton $i$ and the measurement of the observable $x_{\rm trk}$.
$\mathcal{H}$ is the inclusive hard function, which contains the PDFs and the perturbative cross section for producing a parton $i$ with transverse momentum $p_T/z$ and rapidity $\eta$.
The argument $z$ of $\mathcal{G}$ denotes the momentum fraction of the jet with respect to the initial parton produced by the hard function $\mathcal{H}_i$, such that the jet transverse momentum is $p_T/z \times z = p_T$. A sum over the repeated flavor index $i$ is understood and the convolution structure over the momentum fraction $z$ is explicitly given as 
\begin{align}
\label{eq:convol}
 [\mathcal{H}_{i}\otimes_z  \mathcal{G}_{i\to \rm{trk}}] (p_T, \eta, x_{\rm trk},\mu)&=\sum_i\! \int_0^\infty\!\! \df p_T' \! \int_0^1\! \df z\, \mathcal{H}_{i}(p_T',\eta, \mu)\, \mathcal{G}_{i\to {\rm trk}}(z,x_{\rm trk},p_T R,\mu) \delta(p_T \!-\! z p'_T)\nn \\
& =\sum_i \int_0^1 \frac{\mathrm{d} z}{z}\, \mathcal{H}_{i}(p_{T}/z,\eta, \mu)\, \mathcal{G}_{i\to {\rm trk}}(z,x_{\rm trk},p_T R,\mu)\,.
\end{align}
The jet function $\mathcal{G}$ is a multiscale function, describing both the jet dynamics at the scale $p_T R$, as well as the non-perturbative dynamics at the scale $\Lambda_{\text{QCD}}$, through the dependence on $x_{\rm trk}$. It also depends explicitly on the jet algorithm used to define the jets.

 As compared to a measurement of a single hadron fragmentation function in a jet~\cite{Kang:2016ehg}, the measurement of the track energy fraction does not modify the proof of the factorization \cite{Collins:1981ta,Bodwin:1984hc,Collins:1985ue,Collins:1988ig,Collins:1989gx,Nayak:2005rt,Collins:2011zzd}. It only modifies the explicit form of the jet function appearing in \Eq{eq:convol}. Much like for the extraction of universal PDFs and FFs, our ability to extract universal field-theoretic matrix elements relies crucially on rigorous factorization theorems in the hadron collider environment, highlighting the seminal work of  \cite{Collins:1989gx,Collins:1988ig,Collins:1985ue}, and emphasizing the importance of further understanding factorization and its breaking in hadron-hadron collisions (see e.g. \cite{Collins:2007nk,Collins:2007jp,Bomhof:2007su,Rogers:2010dm,Buffing:2013dxa,Gaunt:2014ska,Zeng:2015iba,Catani:2011st,Schwartz:2017nmr,Forshaw:2008cq,Forshaw:2006fk,Martinez:2018ffw,Angeles-Martinez:2016dph,Forshaw:2012bi,Angeles-Martinez:2015rna,Schwartz:2018obd,Rothstein:2016bsq,Forshaw:2021fxs} for recent progress).

The track jet functions can be given a field theoretic definition in Soft-Collinear Effective Theory (SCET) \cite{Bauer:2000ew, Bauer:2000yr, Bauer:2001ct, Bauer:2001yt,Rothstein:2016bsq}, following the treatment of fragmenting jet functions. We define the quark and gluon track jet functions as 
\begin{align}
\mathcal{G}_{q\to \rm{trk}}(z,x_{\rm trk},p_T R,\mu)&= \frac{z}{2N_c}\,  \delta \biggl( x_{\text{trk}}\!-\!\frac{P_R^-}{2p_T \cosh \eta}\biggr) \nn \\
&\quad \times \sum_{J,X} \text{tr} \biggl[\frac{\bnslash}{2} \langle 0 |\delta (\omega- \bar n \cdot \mathcal P) \chi_n(0) |J X \rangle \langle J X |\bar \chi_n(0)|0 \rangle  \biggr]\,, \\
\mathcal{G}_{g\to \rm{trk}}(z,x_{\rm trk},p_T R,\mu)&= -\frac{z \omega}{(d-2)(N_c^2-1)}\,   \delta \biggl( x_{\text{trk}}\!-\!\frac{P_R^-}{2p_T\cosh(\eta)}\biggr)    \nn \\
&\quad \times \sum_{J,X} \langle 0 | \delta(\omega- \bar n \cdot \mathcal P ) \mathcal{B}_{n \perp \mu}(0) | J X \rangle \langle J X| \mathcal B_{n\perp}^\mu (0) |0 \rangle\,.
\end{align}
Here $J$ denotes the identified jet state with large light-cone momentum $p_T$,  as defined using an appropriate jet algorithm with jet radius $R$. We sum over all unidentified states $X$, and all jet states, $J$, consistent with the jet definition. $P_R^-$ denotes the large light-cone momentum of the charged hadrons in the jet. The matrix elements are averaged over color and spin, and we have used $d$ space-time dimensions as a regularization. The fields $\chi$ and $\mathcal{B}$ are the collinear quark and gluon fields of SCET \cite{Bauer:2000ew, Bauer:2000yr, Bauer:2001ct, Bauer:2001yt,Rothstein:2016bsq}. The variable $\omega$ denotes the large light-cone momentum component of the field initiating the jet. The track jet function is expressed in terms of the variable 
\begin{align}
z=\frac{p_T}{\omega}\,.
\end{align}
The resulting track jet function ends up being a function of the combination $p_T R$.

In the case that $p_T R \gg \Lambda_{\text{QCD}}$, we can factorize $\mathcal{G}$ into a convolution of a perturbative matching coefficient and non-perturbative matrix element(s). This factorization can be efficiently achieved using SCET, and the track functions onto which we will match are defined in terms of the SCET fields as  \cite{Chang:2013rca,Chang:2013iba} 
\begin{align}
T_q(x,\mu)&= \frac{16 \pi^3}{2N_c}  \sum_X \delta \biggl( x\!-\!\frac{P_R^-}{\omega}\biggr) \text{tr} \biggl[\frac{\bnslash}{2} \langle 0 |\delta (\omega- \bar n \cdot \mathcal P) \chi_n(0) |X \rangle \langle X |\bar \chi_n(0)|0 \rangle  \biggr]\,, \\
T_g(x,\mu)&= -\frac{16 \pi^3 \omega}{(d-2)(N_c^2-1)} \,      \nn \\
&\quad \times \sum_X \delta \biggl( x\!-\!\frac{P_R^-}{\omega}\biggr) \langle 0 | \delta(\omega- \bar n \cdot \mathcal P )\delta^2(\mathcal{P}_\perp) \mathcal{B}_{n \perp \mu}(0) | X \rangle \langle X| \mathcal B_{n\perp}^\mu (0) |0 \rangle\,.
\end{align}
As compared to the track jet function, there is no identified jet, and hence no jet algorithm dependence, making the track function a universal non-perturbative function of QCD. The matching of the track jet function $\mathcal{G}$ onto products of these universal track functions $T_i(x)$ takes the following form
\begin{align}
\label{eq:G_match}
\mathcal{G}_{i\to \rm{trk}}(z,x_{\rm trk},p_T R,\mu) =& \sum_{m=1}\,\sum_{i_1,.,i_m}\,\int \prod_{k=1}^m\bigl[ \df y_k\,\df x_k T_{i_k}(x_k)\bigr]\,\delta\biggl(x_{\rm trk}-\sum_{k=1}^mx_k y_k\biggr)\,\delta\biggl(\sum_{k=1}^my_k - 1\biggr)\nonumber\\
&\times\, \mathcal{J}_{i\to[i_1,...,i_m]}(z,y_1,...,y_{m-1},p_T R,\mu) +\mathcal{O}\biggl (\frac{\Lambda_{\text{QCD}}}{p_T R} \biggr)\,\nn\\
\equiv&\sum_{m=1}\,\sum_{i_1,.,i_m}\, \mathcal{J}_{i\to[i_1,...,i_m]} \otimes  \prod_{k=1}^m T_{i_k}(x_{\rm trk}) +\mathcal{O} \biggl (\frac{\Lambda_{\text{QCD}}}{p_T R} \biggr)\,,
\end{align}
where the definition of the convolution structure involving different number of track functions can be inferred by comparing the two lines. Since the measurement of the track energy fraction in a jet requires one to track all the hadrons (as compared to fragmentation), the matching coefficients are differential in an arbitrary number of final state hadrons. At $\mathcal{O}(\alpha_s^N)$ in perturbation the maximum number of partons is $N+1$. In \Eq{eq:G_match} we have also used that $\mathcal{J}_{i\to[i_1,...,i_m]}$ can always be written independent of the momentum fraction $y_m$ by using momentum conservation, expressed explicitly by the delta function $\delta(\sum_{k=1}^my_k - 1)$. 

\subsection{Renormalization Group Evolution}\label{sec:RG}

The factorization theorem in \Eq{eq:fact} guarantees that the function $\mathcal{G}$ evolves with the timelike DGLAP evolution, namely 
\begin{align}\label{eq:dglap_conv}
\frac{\mathrm{d}}{\mathrm{d} \ln\mu^2}\, \mathcal{G}_{i\to \rm{trk}}(z,x_{\rm trk},p_T R,\mu)&= \sum_j\int\limits_z^1 \frac{\df z'}{z'}\, P_{ji}\Bigl ( \frac{z}{z'}  \Bigr)\, \mathcal{G}_{j\to \rm{trk}}(z',x_{\rm trk},p_T R,\mu)\,.
\end{align}
We follow the convention for the normalization of the timelike splitting functions used in \cite{Mitov:2006wy,Mitov:2006ic,Mitov:2006xs}, and expand it in terms of the reduced coupling $a_s =\frac{\alpha_s}{4\pi}$ as
\begin{align}
P_{ij}(z)=\sum_{L=0}^\infty a_s^{L+1}P_{ij}^{(L)}(z)
\,.\end{align}
The timelike splitting function is known to NNLO \cite{Stratmann:1996hn,Mitov:2006ic,Moch:2007tx,Almasy:2011eq,Chen:2020uvt}.  The evolution of the track jet function can then similarly be expanded as 
\begin{align}
\frac{\mathrm{d}}{\mathrm{d} \ln\mu^2}\, \mathcal{G}_{i\to \rm{trk}}(z,x_{\rm trk},p_T R,\mu)&=a_s P^{(0)}_{ji} \otimes_z  \mathcal{G}_{j\to \rm{trk}}(z,x_{\rm trk},p_T R,\mu) \nonumber\\
&\quad + a_s^2 P^{(1)}_{ji} \otimes_z  \mathcal{G}_{j\to \rm{trk}}(z,x_{\rm trk},p_T R,\mu) + \cdots\,,
\end{align}
where we have introduced the shorthand notation $\otimes_z$ to express the convolution defined in eq.~\eqref{eq:dglap_conv}. 

The RG equations for the track function are also known, but they exhibit a more complicated non-linear evolution equation \cite{Chang:2013rca,Chang:2013iba,Chen:2022pdu,Chen:2022muj,Jaarsma:2022kdd,Li:2021zcf}, due to the fact that they measure the energy fraction on \emph{all} charged hadrons, as opposed to a single identified hadron. Their evolution is described by perturbative kernels, $K_{i\to i_1 \cdots i_k}$, that capture the mixing between a single track function $T_i$ and a product of $k$ track functions, $T_{i_1} \cdots  T_{i_k}$.
To NLO, or $\mathcal{O}(\alpha_s^2)$, their evolution equations are given by
\begin{align}\label{eq:track_evolution}
\frac{\mathrm{d}}{\mathrm{d} \ln \mu^2}\, T_i(x)=a_s & {\Bigl[K_{i \rightarrow i}^{(0)} T_i+K_{i \rightarrow i_1 i_2}^{(0)} \otimes T_{i_1} T_{i_2}\Bigr](x) } \nonumber\\
& +a_s^2\Bigl[K_{i \rightarrow i}^{(1)} T_i+K_{i \rightarrow i_1 i_2}^{(1)} \otimes T_{i_1} T_{i_2}+K_{i \rightarrow i_1 i_2 i_3}^{(1)} \otimes T_{i_1} T_{i_2} T_{i_3}\Bigr](x)\,,
\end{align}
where the explicit expressions of these multi-parton final state kernels are given to NLO for all channels in ref.~\cite{Chen:2022pdu}. The convolution structures for the $1\to2$ and $1\to 3$ kernels are defined as
\begin{align}
    &K_{i\rightarrow i_1 i_2} \otimes T_{i_1} T_{i_2} (x)
    \\
    &=
    \int_0^1 \df x_1 \df x_2 \ T_{i_1}(x_1) T_{i_2}(x_2)
    \int_0^1 \df z_1 \df z_2 \ 
    \delta(1\!-\!z_1\!-\!z_2) \delta(x\!-\!z_1x_1\!-\!z_2x_2) 
    K_{i\rightarrow i_1 i_2}(z_1,z_2)\,,
    \nn
    \\
    &K_{i\rightarrow i_1 i_2 i_3} \otimes T_{i_1} T_{i_2} T_{i_3} (x)
    \nn \\
    &=
    \int_0^1 \df x_1 \df x_2 \df x_3 \ 
    T_{i_1}(x_1) T_{i_2}(x_2) T_{i_3}(x_3)
    \nn
    \\
    &\qquad\times
    \int_0^1 \df z_1 \df z_2 \df z_3 \  
    \delta(1\!-\!z_1\!-\!z_2\!-\!z_3) 
    \delta(x\!-\!z_1x_1\!-\!z_2x_2\!-\!z_3x_3) 
    K_{i\rightarrow i_1 i_2 i_3}(z_1,z_2,z_3)\,.
\end{align}
We note that the timelike DGLAP splitting kernels can in fact be written in terms of the  kernels for the track function evolution \cite{Chen:2022pdu}
\begin{align} \label{eq:split}
P^{(0)}_{ji}(z)\equiv P^{(0)}_{i\to j}(z) &= \delta_{ij}K_{i\to i}^{(0)} + \sum_k K_{i\to jk}^{(0)}(z,1-z)\,,\\
P^{(1)}_{ji}(z)\equiv P^{(1)}_{i\to j}(z) &= \delta_{ij}K_{i\to i}^{(1)} + \sum_k K_{i\to jk}^{(1)}(z,1-z) \nonumber\\
&\quad + \sum_{k,l} \int\! \df z'\, \df z{''}\, \delta(1-z-z'-z'')\, K_{i\to jkl}^{(1)}(z,z',z'')\,.
\end{align}
These kernels provide a unified description of collinear evolution equations.

Renormalization group consistency fixes the RG evolution of the perturbative matching kernels $\mathcal{J}$ in \Eq{eq:G_match} to be the difference of the DGLAP and the track function evolution, as is standard for such refactorizations.
Hence they also evolve with a complicated non-linear RG equation. This is of course expected, since $\mathcal{G}$ is sensitive to the charge fraction of all hadrons within a jet. However, the RG evolution is completely fixed, and is known from recent calculations of the track function RG \cite{Chen:2022pdu,Chen:2022muj,Jaarsma:2022kdd,Li:2021zcf}.

\subsection{Moments}\label{sec:mom}

Moments of the track functions play a key role since they appear in factorization theorems for energy correlators \cite{Chen:2020vvp}, and exhibit simpler evolution equations \cite{Jaarsma:2022kdd,Li:2021zcf}. We define the moments of a track function of flavor $i$ as
\begin{align} \label{eq:T_mom}
T_i(n,\mu)=\int_0^1 \df x~ x^n~ T_i (x,\mu)\,.
\end{align}
They obey the sum rule
\begin{align}\label{T0}
T_i(0,\mu)=1\,,
\end{align}
corresponding to probability conservation.
Since moments of the track jet function can also be directly measured experimentally, it is important to understand the relation between moments of the track jet function, and moments of the track functions itself.

Taking moments of eq.~\eqref{eq:G_match} with respect to the track energy fraction, we find
\begin{align}
\frac{\mathrm{d} \sigma(N)}{\mathrm{d} p_{T}\,\mathrm{d} \eta} \equiv \int \df x_{\rm trk}\, x_{\rm trk}^N\,\frac{\mathrm{d} \sigma}{\mathrm{d} p_{T}\,\mathrm{d} \eta\, \mathrm{d} x_{\rm trk}}= \mathcal{H}_{i}(p_{T}/z,\eta, \mu)\otimes_z  \mathcal{G}_{i\to \rm{trk}}(z,N,p_T R,\mu)\,,
\end{align}
where
\begin{align}
\label{eq:G}
\mathcal{G}_{i\to \rm{trk}}(z,N,p_T R,\mu) &= \sum_{m=1}^{\infty}\,\sum_{i_1,.,i_m}\,\sum_{\substack{\sum_{i}^m n_i=N;\\n_i\geq 0}}
\left(\begin{array}{c}
N \\
n_1, n_2, \ldots, n_m
\end{array}\right)\, \int \prod_{k=1}^m \bigl[  \df x_k\, x_k^{n_k}\,T_{i_k}(x_k) \bigr] \nonumber\\
&\times\int \prod_{k=1}^m \bigl[ \df y_k\, y_k^{n_k} \bigr]\mathcal{J}_{i\to[i_1,...,i_m]}(z,y_1,...,y_{n-1},p_T R,\mu)\,\delta\biggl(\sum_{k=1}^my_k - 1\biggr) \nnu
\equiv& \sum_{m=1}^{\infty}\,\sum_{i_1,.,i_m}\,\sum_{\substack{\sum_{i}^m n_i=N;\\n_i\geq 0}}
\left(\begin{array}{c}
N \\
n_1, n_2, \ldots, n_m
\end{array}\right)\,\prod_{k=1}^m T_{i_k}(n_k)\nnu
&\times\, \mathcal{J}_{i\to[i_1,...,i_m]}(z,n_1,n_2,...,n_{m-1},p_T R,\mu)\,.
\end{align}
Terms involving $T_i(0)$ can be simplified using the sum rule in \Eq{T0}. This provides an explicit relationship between the moments of the universal non-perturbative track functions and the moments of charged energy distribution inside jets at the LHC, and will thus play a key role in the extraction of track functions from experiment. This relation also shows that it is possible to directly observe the renormalization group evolution of the track function moments by measuring the moments of the $x_{\text{trk}}$ distribution. We will discuss this in detail in \Sec{sec:pheno}. Note that the zeroth-moment of $\mathcal{G}_{i\to \text{trk}}$ gives back the semi-inclusive jet function from ref.~\cite{Kang:2016mcy}, due to the sum-rule in \eqref{T0}.  

\subsection{Relation to Multi-Hadron Fragmenting Jet Functions}\label{sec:multi}

Although the main focus of this paper is on developing the formalism necessary to extract track functions from hadron-collider data, we also note that the formalism introduced here allows for the calculation of $N$-hadron fragmentation within jets. This is due to the relation between track functions and multi-hadron fragmentation functions discovered in refs.~\cite{Chen:2022pdu,Chen:2022muj}, that was used to obtain the renormalization group equations of the latter from the former. In our case we can use it to obtain the matching for (multi-hadron) fragmenting jet functions from that of track jet functions. See refs.~\cite{deFlorian:2003cg,Majumder:2004wh,Ceccopieri:2007ip,Cocuzza:2023oam} where di-hadron fragmentation functions were discussed and extracted from the available data.

The key insight of refs.~\cite{Chen:2022pdu,Chen:2022muj} was that since the track functions measure the energy fraction on all hadrons, they must contain in them the renormalization group evolution for $N$-hadron fragmentation functions, for all $N$. In this sense, the RG equations for track functions can be viewed as generating functionals for the RG equations of $N$-hadron fragmentation functions. Intuitively, the relationship is clear: to reduce an equation involving track functions to a similar equation involving $N$-hadron fragmentation functions, one must integrate out all but $N$ hadrons. At a mathematical level, this is achieved by replacing (some of) the track functions $T_i(x)\to \delta(x)$, integrating over the energy fraction, and then relabelling the remaining track functions as (multi-hadron) fragmentation functions. In the general $N$-hadron case, this also involves summing over all ways to divide the set of $N$ hadrons into non-empty sets. The detailed procedure was described in ref.~\cite{Chen:2022pdu}.

Interestingly, we can view the matching from  $\mathcal{G}$ onto products of track functions in the same manner. Since $\mathcal{G}$ tracks the energy fraction of \emph{all} hadrons in the jet, it can be viewed as a generating functional for $N$-hadron fragmenting jet functions. These can be derived in an identical manner as presented in ref.~\cite{Chen:2022pdu} for the renormalization equation.

Applying these rules to the track jet function, we first sum over all ways of replacing track functions by delta functions 
\begin{align}
 &\mathcal{G}_{i\to \rm{trk}}(z,x_{\rm trk},p_T R,\mu)\\
  &\to\sum_{m\geq 1}\, \sum_{ \{i_f\} }\,\sum_{ f_1\in\{1,2,\cdots,m\} }
  \,\mathcal{J}_{i\to[i_1,...,i_m]}\otimes 
  T_{i_{f_1}}(x_{f_1})\delta(x_{f_2})\cdots\delta(x_{f_m})\nn\\
  &+ \sum_{m\geq 2}\, \sum_{ \{i_f\} }\,\sum_{ \{f_1,f_2\}\atop\subseteq\{1,2,\cdots,m\} }
  \,\mathcal{J}_{i\to[i_1,...,i_m]}\otimes 
  T_{i_{f_1}}(x_{f_1})T_{i_{f_2}}(x_{f_2})\delta(x_{f_3})\cdots\delta(x_{f_m})\nn\\
  &+ \sum_{m\geq 3}\, \sum_{ \{i_f\} }\,\sum_{ \{f_1,f_2,f_3\}\atop\subseteq\{1,2,\cdots,m\} }
  \,\mathcal{J}_{i\to[i_1,...,i_m]}\otimes 
  T_{i_{f_1}}(x_{f_1})T_{i_{f_2}}(x_{f_2})T_{i_{f_3}}(x_{f_3})\delta(x_{f_4})\cdots\delta(x_{f_m})\nn\\
  &+\cdots\nn\\
  &+\sum_{m\geq N}\, \sum_{ \{i_f\} }\,\sum_{ \{f_1,f_2,\cdots,f_N\}\atop\subseteq\{1,2,\cdots,m\} }
  \,\mathcal{J}_{i\to[i_1,...,i_m]}\otimes 
  T_{i_{f_1}}(x_{f_1})T_{i_{f_2}}(x_{f_2})\cdots T_{i_{f_N}}(x_{f_N})\delta(x_{f_{N+1}})\cdots\delta(x_{f_m})\nn
\end{align}
We then replace these with (products of) multi-hadron fragmentation functions
\begin{align}
&\mathcal{G}_{i\to h_1 h_2 \cdots h_N}(z,y_1,y_2,...,y_n,p_T R,\mu)\\
  &=\sum_{m\geq 1}\, \sum_{ \{i_f\} }\,\sum_{ f_1\in\{1,2,\cdots,m\} }
  \,\mathcal{J}_{i\to[i_1,...,i_m]}\otimes 
  D_{i_{f_1}\to h_1h_2\cdots h_N}(y'_{f_1,1},y'_{f_1,2},\cdots,y'_{f_1,N})\nn\\
  & + \sum_{m\geq 2}\, \sum_{ \{i_f\} }\sum_{ \{f_1,f_2\}\atop\subseteq\{1,2,\cdots,m\} }
  \,\mathcal{J}_{i\to[i_1,...,i_m]}\otimes 
  \sum_{S_2}
  D_{i_{f_1}\to k\text{ hadrons}}(y'_{f_1,1},\cdots,y'_{f_1,k})\nn\\
  &\qquad\qquad\qquad\qquad\qquad\qquad\qquad\qquad\qquad\quad
  \times D_{i_{f_2}\to (N-k)\text{ hadrons}}(y'_{f_2,k+1},\cdots,y'_{f_2,N})
  \nn\\
  & + \sum_{m\geq 3}\, \sum_{ \{i_f\} }\sum_{ \{f_1,f_2,f_3\}\atop\subseteq\{1,2,\cdots,m\} }
  \,\mathcal{J}_{i\to[i_1,...,i_m]}\otimes 
  \sum_{S_3}
  D_{i_{f_1}\to k_1\text{ hadrons}}(y'_{f_1,1},\cdots,y'_{f_1,k_1})\nn\\
  &\qquad\times
  D_{i_{f_2}\to k_2\text{ hadrons}}(y'_{f_2,k_1+1},\cdots,y'_{f_2,k_1+k_2})
  D_{i_{f_3}\to (n-k_1-k_2)\text{ hadrons}}(y'_{f_3,k_1+k_2+1},\cdots,y'_{f_2,N})\nn\\
  & +\cdots\nn\\
  &+\sum_{m\geq N}\, \sum_{ \{i_f\} }\,\sum_{ \{f_1,f_2,\cdots,f_N\}\atop\subseteq\{1,2,\cdots,m\} }
  \,\mathcal{J}_{i\to[i_1,...,i_m]}\otimes 
  \sum_{S_N}
  D_{i_{f_1}\to 1\text{ hadron}}(y'_{f_1,1})\times\cdots 
  \times D_{i_{f_N}\to 1\text{ hadron}}(y'_{f_N,N})\nn\,,
\end{align} 
where $S_k$ denote all ways of dividing the set of hadrons into $k$ non-empty sets. For details, see ref.~\cite{Chen:2022pdu}. This provides an expression for the general $N$-hadron fragmenting jet function.

This expression is useful in two ways. First, as a check on our results, we have used this procedure to derive the (semi-inclusive) single-hadron fragmenting jet function, which was computed to $\mathcal{O}(\alpha_s)$ in ref.~\cite{Kang:2016ehg}. Second, it provides a practical means of computing the (semi-inclusive) multi-hadron fragmenting jet functions. It would be interesting to investigate phenomenological applications of these functions in future work. In particular, this formalism could be useful for the extraction of multi-hadron fragmentation functions from jet measurements at the LHC.

\section{Perturbative Calculations of Matching Coefficients}\label{sec:pert}

We now perform the perturbative calculation of the matching coefficients $\mathcal{J}$ between the track jet function and the universal track functions. We consider the track jet functions up to $\mathcal{O}(\alpha_s^2)$. At this order, we can have at most three particles in the final state, and thus eq.~\eqref{eq:G_match} simplifies to 
\begin{align}
\label{eq:GNNLO}
&\mathcal{G}_{i\to \rm{trk}}(z,x_{\rm trk},p_T R,\mu) = 
 \\ &\quad  \mathcal{J}_{i\to[i]}^{(0)} T_i(x_{\rm trk}) +a_s^1\,\biggl[\mathcal{J}^{(1)}_{i\to[j]}\,T_j(x_{\rm trk})+ \mathcal{J}^{(1)}_{i\to[j,k]}\otimes\,T_j T_k(x_{\rm trk})\biggr]\nonumber\\
&\quad +a_s^2\,\bigl[\mathcal{J}^{(2)}_{i\to[j]}\otimes\,T_j(x_{\rm trk})+ \mathcal{J}^{(2)}_{i\to[j,k]}\otimes\,T_j T_k(x_{\rm trk})
+ \mathcal{J}^{(2)}_{i\to[j,k,l]}\otimes\,T_j T_k T_l(x_{\rm trk})\bigr]\,,
\nn \end{align}
where a sum over the repeated indices is implied.  Since all $m$ partons in the matching coefficients are required to be inside the jet in order to contribute to the track momentum fraction, the jet momentum fraction $z = 1$ for the first order at which this happens, $\mathcal{J}^{(m-1)}_{i\to[i_1,...,i_m]} \propto \delta(1-z)$. 

Using the fact that both the timelike DGLAP and the track function evolution generate single-logarithmic series, we can parameterize the matching coefficients as
\begin{align}
\label{eq:matching1}
\mathcal{J}_{i\to[j]}^{(0)}(z)  &=\delta_{ij}\delta(1-z)\,,\\
\mathcal{J}_{i\to[j]}^{(1)}(z,p_TR,\mu) &= \delta(1-z)\bigl[A_{i\to j}^{(1,1)} L + A_{i\to j}^{(1,0)}\bigl] + B_{i\to j}^{(1,1)}(z) L + B^{(1,0)}_{i\to j}(z) \,,\\
\mathcal{J}_{i\to[j]}^{(2)}(z,p_TR,\mu) &= \delta(1-z)\bigl[\textcolor{blue}{A_{i\to j}^{(2,2)}} L^2 + \textcolor{blue}{A_{i\to j}^{(2,1)}} L + \textcolor{red}{A_{i\to j}^{(2,0)}} \bigl] \nonumber\\
&\quad + \textcolor{blue}{B_{i\to j}^{(2,2)}(z)} L^2 + \textcolor{blue}{B^{(2,1)}_{i\to j}(z)} L+ \textcolor{red}{B^{(2,0)}_{i\to j}(z)} \,,\\
\mathcal{J}_{i\to[j,k]}^{(1)}(z,y_1,p_TR,\mu) &= \delta(1-z)\bigl[C_{i\to jk}^{{(1,1)}}(y_1) L + C_{i\to jk}^{(1,0)}(y_1)\bigl]\,,\\
\mathcal{J}_{i\to[j,k]}^{(2)}(z,y_1,p_TR,\mu) &= \delta(1-z)\bigl[\textcolor{blue}{C_{i\to jk}^{(2,2)}(y_1)} L^2 + \textcolor{blue}{C_{i\to jk}^{(2,1)}(y_1)} L + \textcolor{red}{C_{i\to jk}^{(2,0)}(y_1)} \bigl] \nonumber\\
&\quad + \textcolor{blue}{D_{i\to jk}^{(2,2)}(z,y_1)} L^2 + \textcolor{blue}{D^{{(2,1)}}_{i\to jk}(z,y_1)} L+ \textcolor{red}{D^{(2,0)}_{i\to jk}(z,y_1)} \,, \\
\mathcal{J}_{i\to[j,k,l]}^{(2)}(z,y_1,y_2,p_TR,\mu) &= \delta(1-z) \label{eq:matching2} \\
&\quad \times \bigl[\textcolor{blue}{F_{i\to jkl}^{(2,2)}(y_1,y_2)} L^2 + \textcolor{blue}{F_{i\to jkl}^{(2,1)}(y_1,y_2)} L + \textcolor{red}{F_{i\to jkl}^{(2,0)}(y_1,y_2)} \bigl]\,, 
\nn 
\end{align}
where $L = \ln (\mu^2/(p_T^2 R^2))$. We have separated terms by the order of the logarithms, and by whether they are explicitly proportional to $\delta(1-z)$.

The $1$-loop coefficients can be computed straightforwardly. For $k_T$-type algorithms \cite{Ellis:1993tq,Cacciari:2008gp}, we find
\small
\begin{align}
\label{eq:nloresult}
{\cal J}^{(1)}_{q\to [q,g]}(z,y_q,p_T R,\mu) &= \delta(1-z)\bigg[-L\,P^{(0)}_{qq}(y_q) +4C_F(1+y_q^2)\biggl[\f{\ln(1-y_q)}{1-y_q}\biggr]_+\nn \\
& \hspace{12ex}+ 2C_F(1-y_q)+2 P^{(0)}_{qq}(y_q)\,\ln y_q\bigg]\,,
\nnu
{\cal J}^{(1)}_{q\to [g,q]}(z,y_g,p_T R,\mu) &= {\cal J}^{(1)}_{q\to [q,g]}(z,1-y_q,p_T R,\mu)\,,
\nnu
{\cal J}^{(1)}_{g\to [q,\bar{q}]}(z,y_q,p_T R,\mu) &= \delta(1-z)\bigl[-L\, P^{(0)}_{qg}(y_q)+2P^{(0)}_{qg}(y_q)\ln [y_q(1-y_q)]+4T_F y_q(1-y_q)\bigr]\,, \nnu
{\cal J}^{(1)}_{g\to [\bar{q},q]} (z,y_{\bar{q}},p_T R,\mu) &= {\cal J}^{(1)}_{g\to [q,\bar{q}]}(z,1-y_q,p_T R,\mu)\,,\nnu
{\cal J}^{(1)}_{g\to [g,g]}(z,y_g,p_T R,\mu) &= \delta(1-z)\bigg\{-L  \biggl[C_A\biggl(\frac{4y_g}{(1-y_g)_+}+4(1-y_g)
\left[\frac{1}{y_g}\right]_++4(1-y_g)y_g\biggr)  \nn \\
&\hspace{-2cm} + \beta_0\bigl[\delta(1-y_g)+\delta(y_g)\bigr]\biggr]
 -C_A\biggl(-\frac{8 y_g}{(1-y_g)_+} \ln y_g - 8(1-y_g)\biggl[\frac{\ln y_g}{y_g}\biggr]_+
\nn \\ 
& \hspace{-2cm} -8 (1-y_g)\left[\frac{1}{y_g}\right]_+ \ln (1-y_g) - 8y_g\biggl[\frac{\ln (1-y_g)}{1-y_g}\biggr]_+ - 8y_g(1-y_g)\ln [y_g(1-y_g)]\bigg)\bigg\}\,,
\nnu
{\cal J}^{(1)}_{q\to [q]}(z,p_T R,\mu) &= L\,P^{(0)}_{qq}(z) -4C_F(1+z^2)\biggl[\f{\ln(1-z)}{1-z}\biggr]_+ - 2C_F(1-z) \, ,\nnu
{\cal J}^{(1)}_{q\to [g]}(z,p_T R,\mu)  &=  L\,P^{(0)}_{gq}(z) -2P^{(0)}_{gq}(z)\ln(1-z)-2C_F z \, , \nnu
{\cal J}^{(1)}_{g\to [q]}(z,p_T R,\mu)  &= L\,P^{(0)}_{qg}(z) -2P^{(0)}_{qg}(z)\ln(1-z)-4T_F z(1-z) \, ,\nnu
{\cal J}^{(1)}_{g\to [g]}(z,p_T R,\mu) &= L\,P^{(0)}_{gg}(z)-\,  8C_A\f{(1-z+z^2)^2}{z}\biggl[\f{\ln(1-z)}{1-z}\biggr]_+ \,.
\end{align}
\normalsize
Here the $[~]_+$ denotes a standard plus distribution.
From these expressions, we are able to read off the coefficients for the parameterization in eqs.~\eqref{eq:matching1}-\eqref{eq:matching2} above. Note also that some terms that are proportional to $\delta(1-z)$ are hidden in the DGLAP splitting kernels.

Unlike the $1$-loop computation, the $2$-loop computation of the matching coefficients is much more challenging. The logarithmically-enhanced terms (shown in blue in eqs.~\eqref{eq:matching1}-\eqref{eq:matching2}) are fixed by the renormalization group equations, and will be determined below. The remaining terms (shown in red) can be calculated by combining the approaches developed in refs.~\cite{Ritzmann:2014mka,Liu:2021xzi} and are left to future work. 
The non-trivial ingredient is the NLO evolution of the track functions in \Eq{eq:track_evolution}, which was computed recently in ref.~\cite{Chen:2022pdu}. Performing the perturbative evolution, we are then able to express the blue terms in eqs.~\eqref{eq:matching1}-\eqref{eq:matching2} in terms of the known evolution kernels, and the $\beta$ function, which we expand as
\begin{align}
\beta(\alpha_s)= -2 \alpha_s \sum\limits_{n=0}^\infty \beta_n
\Bigl(\frac{\alpha_s}{4\pi}\Bigr)^{n+1}
\,.
\end{align}
We find 
\begin{align}
A_{i\to j}^{(2,2)} =&\,\frac{1}{2}\delta_{ij}K_{i\to i}^{(0)} \Bigl( K_{i \to i}^{(0)} - \beta_0\Bigr)\,,\\
A_{i\to j}^{(2,1)} =&-\delta_{ij}K_{i\to i}^{(1)} - A_{i\to j}^{(1,0)} \bigl(K_{j\to j}^{(0)} - \beta_0\bigr)\,,\\
B_{i\to j}^{(2,2)}(z) =&-P_{i\to j}^{(0)}(z) K_{j\to j}^{(0)} + \frac{\beta_0}{2}P_{i\to j}^{(0)}(z) + \frac{1}{2}P_{i\to k}^{(0)}\otimes P_{k\to j}^{(0)}(z)\,,\\
B_{i\to j}^{(2,1)}(z) =&P_{i\to j}^{(1)}(z)  -B_{i\to j}^{(1,0)}(z) K_{j\to j}^{(0)} + \beta_0 B_{i\to j}^{(1,0)}(z) + P_{i\to k}^{(0)}(z) A_{k\to j}^{(1,0)} \nn \\
&+ P_{i\to k}^{(0)}\otimes B_{k\to j}^{(1,0)}(z)\,,\\
C_{i\to jk}^{(2,2)}(y_j) =& \frac{1}{2}K_{i\to jk}^{(0)}(y_j,1-y_j)\Bigl(K_{i\to i}^{(0)}+ K_{j\to j}^{(0)} + K_{k\to k}^{(0)} - \beta_0\Bigr)\,,\\
C_{i\to jk}^{(2,1)}(y_j) =&-A_{i\to m}^{(1,0)} K_{m\to jk}^{(0)}(y_j,1-y_j) -  K_{i\to jk}^{(1)}(y_j,1-y_j)\nn \\
&+C_{i\to jk}^{(1,0)}(y_j)\bigl(-K_{j\to j}^{(0)} - K_{k\to k}^{(0)} + \beta_0\bigr)\,,\\
D_{i\to jk}^{(2,2)}(z,y_j) =&-P_{i\to m}^{(0)}(z) K_{m\to jk}^{(0)}(y_j,1-y_j) \\
D_{i\to jk}^{(2,1)}(z,y_j) =&P_{i\to m}^{(0)}(z) C_{m\to jk}^{(1,0)}(y_j) -  B_{i\to m}^{(1,0)}(z) K_{m\to jk}^{(0)}(y_j,1-y_j)\,,\\
F_{i\to jkl}^{(2,2)}(y_j,y_k) =& \frac{1}{3} \int \frac{\df y_m}{y_m}\, \df y_l\,  K_{i\to jm}^{(0)}(y_j,y_m)K_{m\to kl}^{(0)}\Bigl(\frac{y_k}{y_m},\frac{y_l}{y_m}\Bigr)\delta(y_m\!-\!y_k\!-\!y_l)\delta(1\!-\!y_j\!-\!y_k\!-\!y_l)\nonumber\\
&\quad + (j \leftrightarrow k)+ (j \leftrightarrow l)
\,,\\
F_{i\to jkl}^{(2,1)}(y_j,y_k) = 
&-\frac{2}{3}\int \frac{\df y_m}{y_m} \df y_l\,  C^{(1,0)}_{i\to jm}(y_j,y_m)K_{m\to kl}^{(0)}\Bigl(\frac{y_k}{y_m},\frac{y_l}{y_m}\Bigr)\delta(y_m\!-\!y_k\!-\!y_l)\delta(1\!-\!y_j\!-\!y_k\!-\!y_l)\nonumber\\
&\quad + (j \leftrightarrow k)+ (j \leftrightarrow l)
- K_{i\to jkl}^{(1)}(y_j,y_k,1-y_j-y_k)\,.
\end{align}
Here we have used the notation  
$P_{i\to j}^{(n)}(z)= P^{(n)}_{ij}(z)$
for convenience. The sum over repeated indices are implied as long as the indices are not one of the incoming or outgoing states. For example, the RHS expression of $A_{i\to j}^{(2,2)}$ do not have sum over $i$ (or $j$), as $i$ (or $j$) is the index of the incoming (outgoing) state of the LHS. Lastly, although the explicit calculation at $1$-loop demonstrates that $A_{i\to j}^{(1,0)}=0$, we present the full expressions derived from the RG structure without assuming this. 
In our presentation, we find it most convenient to express the dependence on the momentum fraction $z$ of the jet  in terms of the usual DGLAP splitting functions, and track-momentum fractions $y_i$ in terms of the general kernels $K$ appearing in the track function evolutions given to NLO in ref.~\cite{Chen:2022pdu}. We do not reproduce these kernels here due to their length. This provides an explicit result for all the logarithmically enhanced terms of the two-loop matching coefficients between the track jet functions and the universal track functions. 

\begin{figure}
\centering
\includegraphics[width=0.7\textwidth]{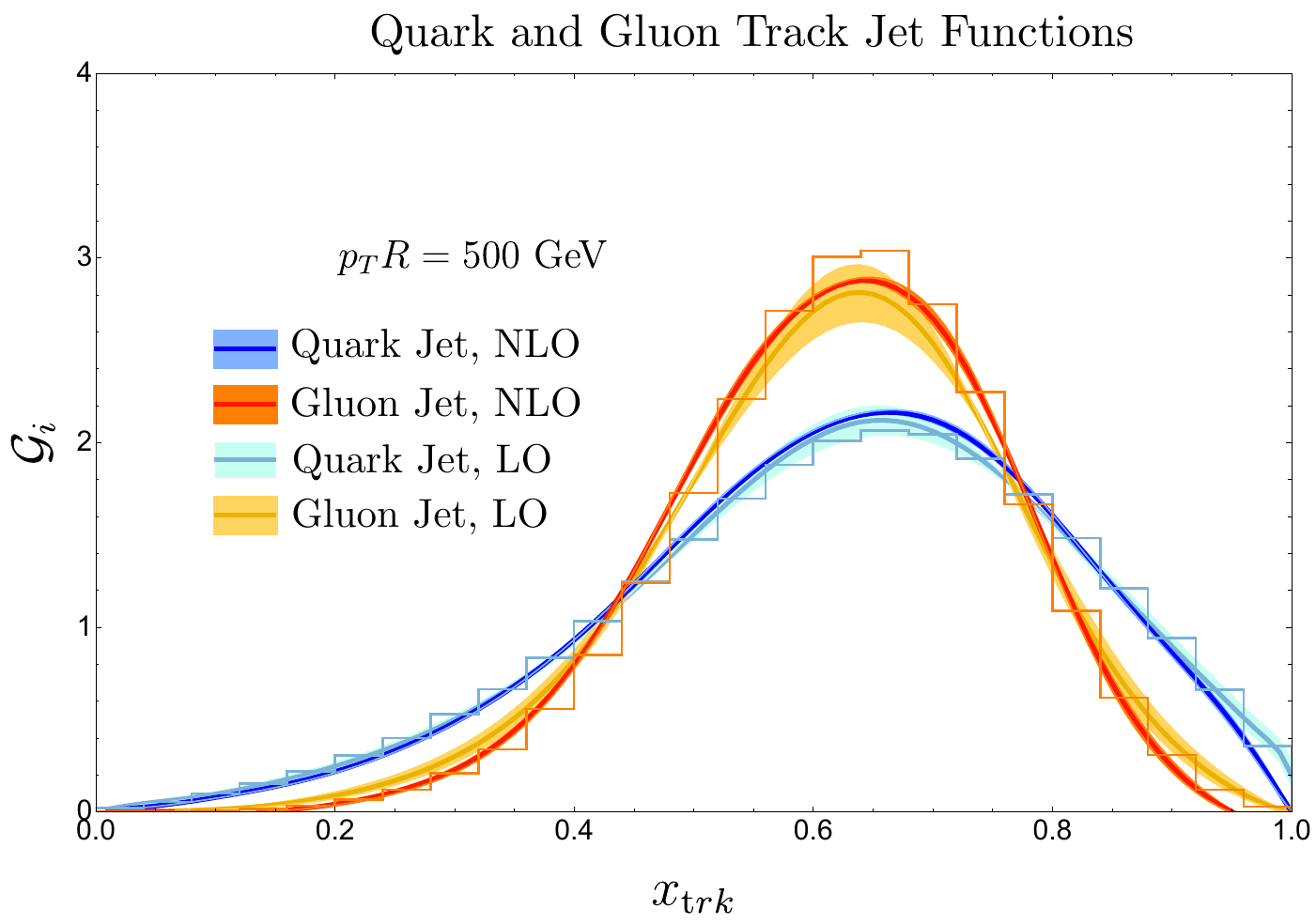}
\caption{The track jet functions at LO and NLO  for both quark and gluon jets. Results of the $x_{\text{trk}}$ distribution from the Pythia parton shower for pure quark and gluons samples (shown as histograms). Good convergence of the perturbative results is observed, as well as good agreement with Pythia. Note that the LO track jet functions are identical to the LO track functions here. As the input LO track functions are extracted using the LO hard coefficients from $e^+e^-$~\cite{Chang:2013rca}, they agree well with Pythia.} \label{fig:track_func}
\end{figure}

\section{Track Jet Functions at Hadron Colliders}\label{sec:pheno}

In this section, we apply our track jet formalism to obtain phenomenological predictions for the track energy fraction in jets in both proton-proton collisions and DIS. We will highlight the impact of the matching coefficients, which appear in the expansion of the track jet functions onto track functions, for extracting the universal track functions from jet measurements. We also feature some properties of the RG evolution of moments of the track energy fraction that would be interesting to measure experimentally.

\begin{figure}
\centering
\includegraphics[width=0.7\textwidth]{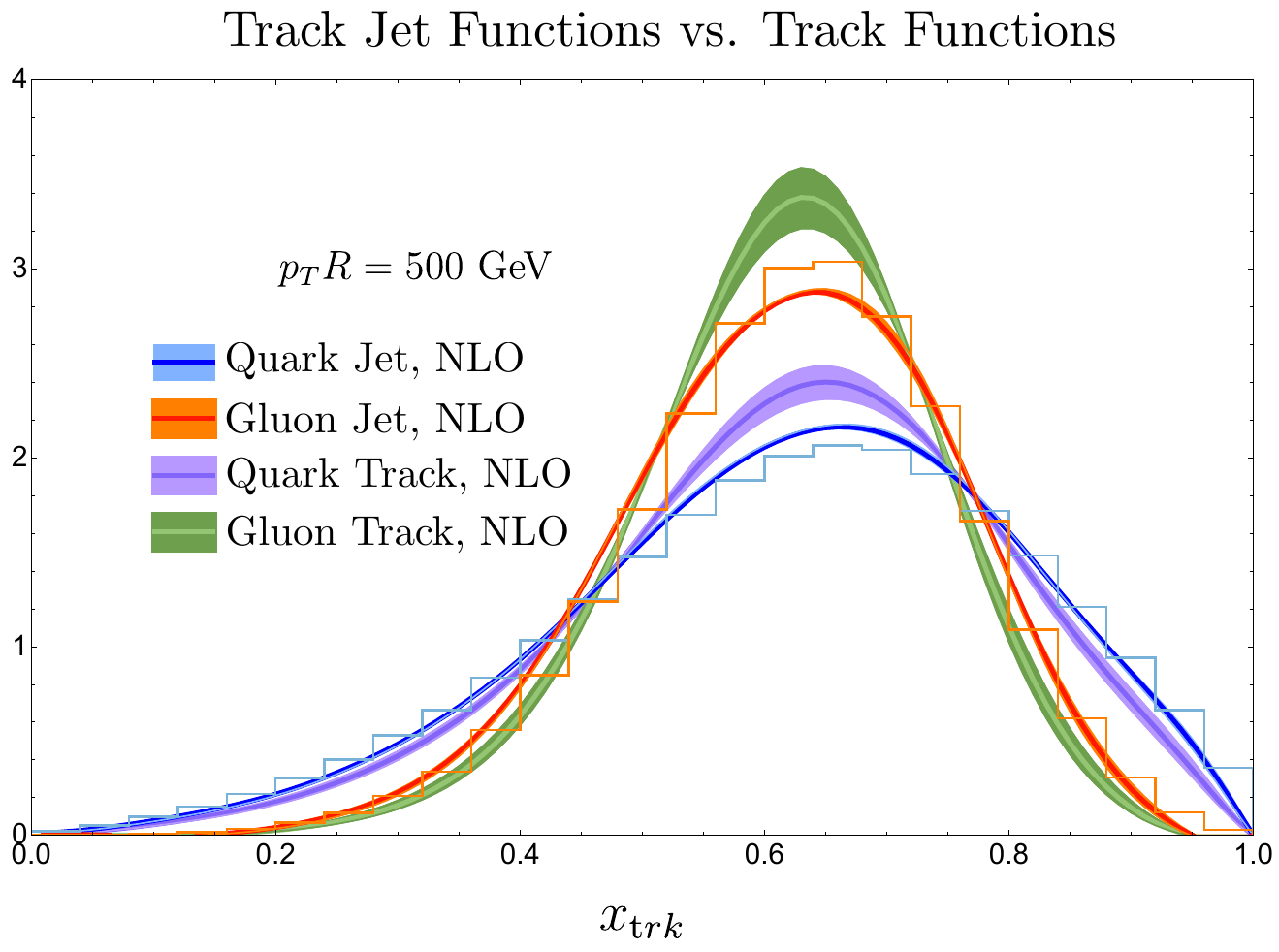}
\caption{A comparison of the NLO track jet function with the pure track function, both evaluated at the scale $\mu=250$ GeV. The corresponding Pythia distribution is also shown (as histogram). The large numerical differences are due to the matching coefficients between the track jet function and the universal track functions, and show the importance of using the track jet function to extract track functions from jet measurements at the LHC.}\label{fig:diff}
\end{figure}

We will use the matching coefficients of the track jet functions at order $\alpha_s^0$ and $\alpha_s$, which are matched with track functions with LL and NLL evolution, respectively. In particular, we will not include the logarithmic terms at order $\alpha_s^2$ that we determined in sec.~\ref{sec:pert}. This choice in accuracy is motivated both by the exceptional convergence of the results for the track energy fraction in $e^+e^-$ \cite{Chen:2022muj}, and by the uncertainties of initial measurements of the track function, which will be limited by the jet energy scale to a few percent \cite{PERF-2011-05,PERF-2016-04,PERF-2015-05,JETM-2018-05}. Furthermore, we find that track functions have smaller scale dependence than their fragmentation functions counterparts, which leads to a reduction in the theoretical uncertainties from scale variation. We use the track functions with an input scale of 100 GeV, extracted in ref.~\cite{Chang:2013rca} and evolve them to the jet scale $ p_T R$ (or $QR$) using eq.~\eqref{eq:track_evolution}, where the matching coefficients are evaluated. The jet scale is varied by a factor $2$ around $p_T R$ (or $QR$) in order to estimate the scale uncertainties.

For full collider predictions, i.e.~without separations into a pure quark or gluon jet sample, we convolve the track jet functions after carrying out NLL DGLAP resummations from the jet scale $\mu_J= p_T R$  (or $Q R$) to the hard scale $\mu_H =p_T$ (or $Q$) with the appropriate process dependent hard functions. For predictions at the LHC we use the NLO hard functions from refs.~\cite{Aversa:1988mm,Aversa:1988fv,Aversa:1988vb,Aversa:1989xw,Aversa:1990uv,Jager:2004jh}, and for predictions in DIS we use the NLO hard functions from ref.~\cite{Anderle:2012rq}. The hard function encapsulate the details of the parton distribution functions and the underlying hard processes, and this evolution resums logarithms of the jet radius $R$, allowing us to account for the quark/gluon fraction of jets. In addition to varying the jet scale, we also vary the hard scale $p_T$ (or $Q$) by a factor of $2$ to estimate the scale uncertainties.

\subsection{LHC}\label{sec:pheno_LHC}

\begin{figure}
\centering
\includegraphics[width=0.7\textwidth]{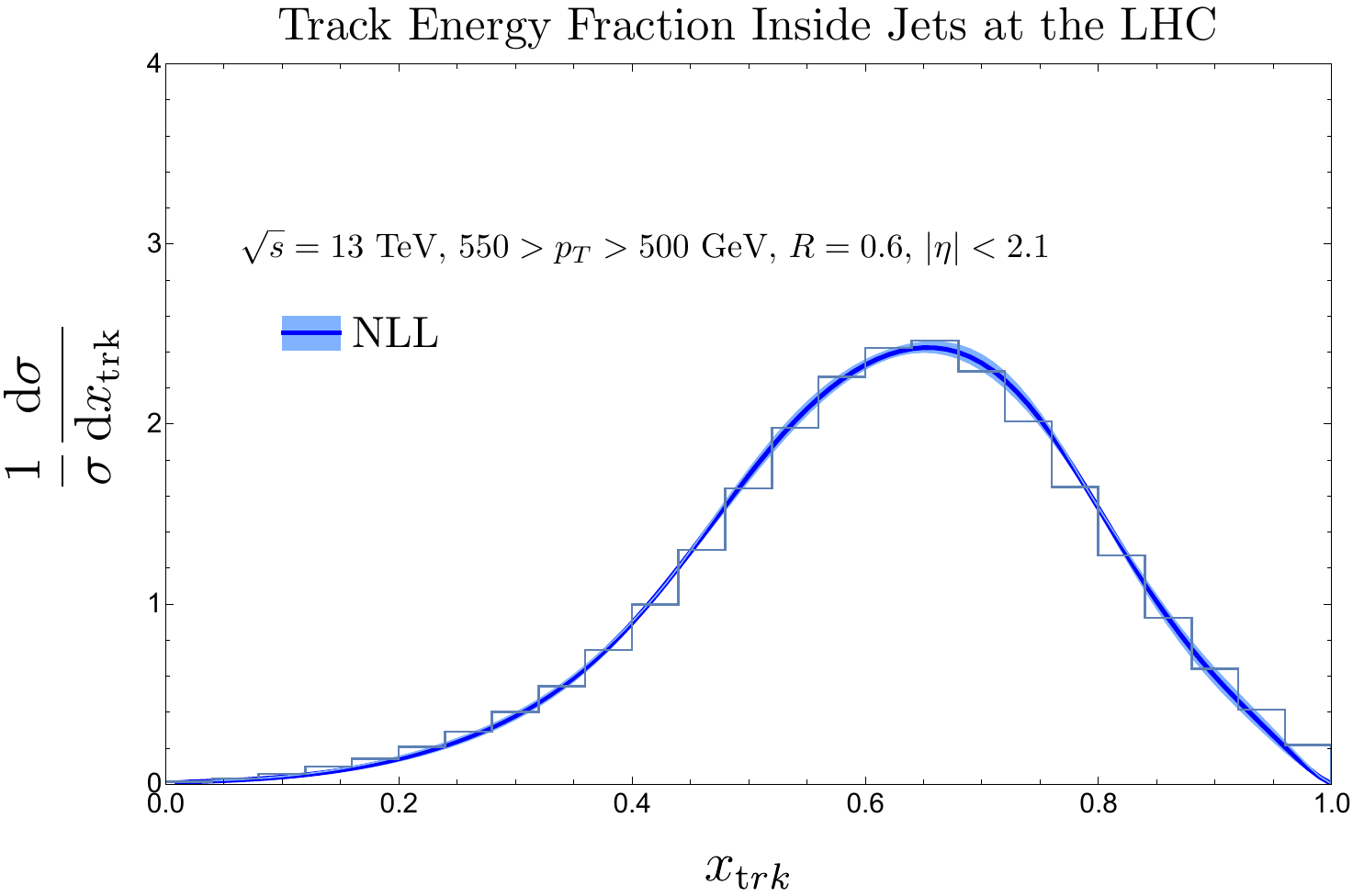}
\caption{The $x_{\text{trk}}$ distribution for $500$ GeV jets at the LHC computed at NLL, and compared with Pythia (shown as histogram). The measurement of this observable will enable the extraction of the universal track functions.} \label{fig:LHC_final}
\end{figure}

We start by considering the case of the LHC, which is of obvious phenomenological interest due to the fact that it is currently operating, it can measure the track functions over a wide range of energies to probe their renormalization group evolution, and it can take advantage of modern state-of-the-art detectors for precision measurements. 

We begin in \Fig{fig:track_func} by plotting the track jet function for anti-$k_T$ quark and gluon jets at the scale $p_T R=500$ GeV (Note that this is not the full $x_\text{trk}$ distribution, which involves the convolution of the quark and gluon track jet functions with the hard function, and is  shown in \Fig{fig:LHC_final}). Results are shown at both LO and NLO\footnote{Meaning LO and NLO matching coefficients were respectively used, as well as LO and NLO evolution kernels to evolve the non-perturbative track functions to the scale of $p_T R$.}, and for both quark and gluon jets. Additionally, we have compared with the Pythia parton shower \cite{Sjostrand:2014zea,Sjostrand:2007gs}. Good convergence of the perturbative series is observed, as well as good agreement with Pythia.

To emphasize the importance of incorporating the perturbative matching coefficients,  we show  in \Fig{fig:diff} a comparison between the universal quark and gluon track functions,  with the  track jet functions. We have evolved both to the scale $\mu=p_T R=500$ GeV to enable a comparison highlighting the impact of the  perturbative matching coefficients. We see a sizeable difference between the two due to matching coefficients mixing quark and gluon track functions contributions at the scale $p_T R$. This illustrates that it is crucial to account for the jet using track jet functions, when extracting track functions from jet measurements at the LHC. 

In \Fig{fig:LHC_final} we show the full physical  $x_\text{trk}$ distribution, which properly incorporates the convolution with the hard function. As compared to \Fig{fig:track_func}, the full distribution further incorporates the resummation of the logarithms from $p_T R$ up to the hard scale $p_T$, and properly accounts for the quark and gluon fractions. The NLL result is compared with Pythia, and excellent agreement is observed. We also see that excellent perturbative accuracy is achieved, certainly sufficient for initial extractions of the track functions from LHC measurements.

An interesting feature of the $x_\text{trk}$ observable, and the track functions themselves, is that they have a simple shape, which should facilitate their extraction from data. This is in contrast to fragmentation functions, $D(z)$, whose distributions cover a large range due to their scaling as $\sim1/z$. Therefore, while the RG evolution of the track functions is much more complicated than for fragmentation functions, the non-perturbative function itself is arguably simpler. Additionally, for precision jet substructure applications, such as to the energy correlators, one is often interested in positive integer moments of the track functions \cite{Chen:2020vvp,Li:2021zcf}, which are even better behaved.

\begin{figure}
\centering
\includegraphics[width=0.75\textwidth]{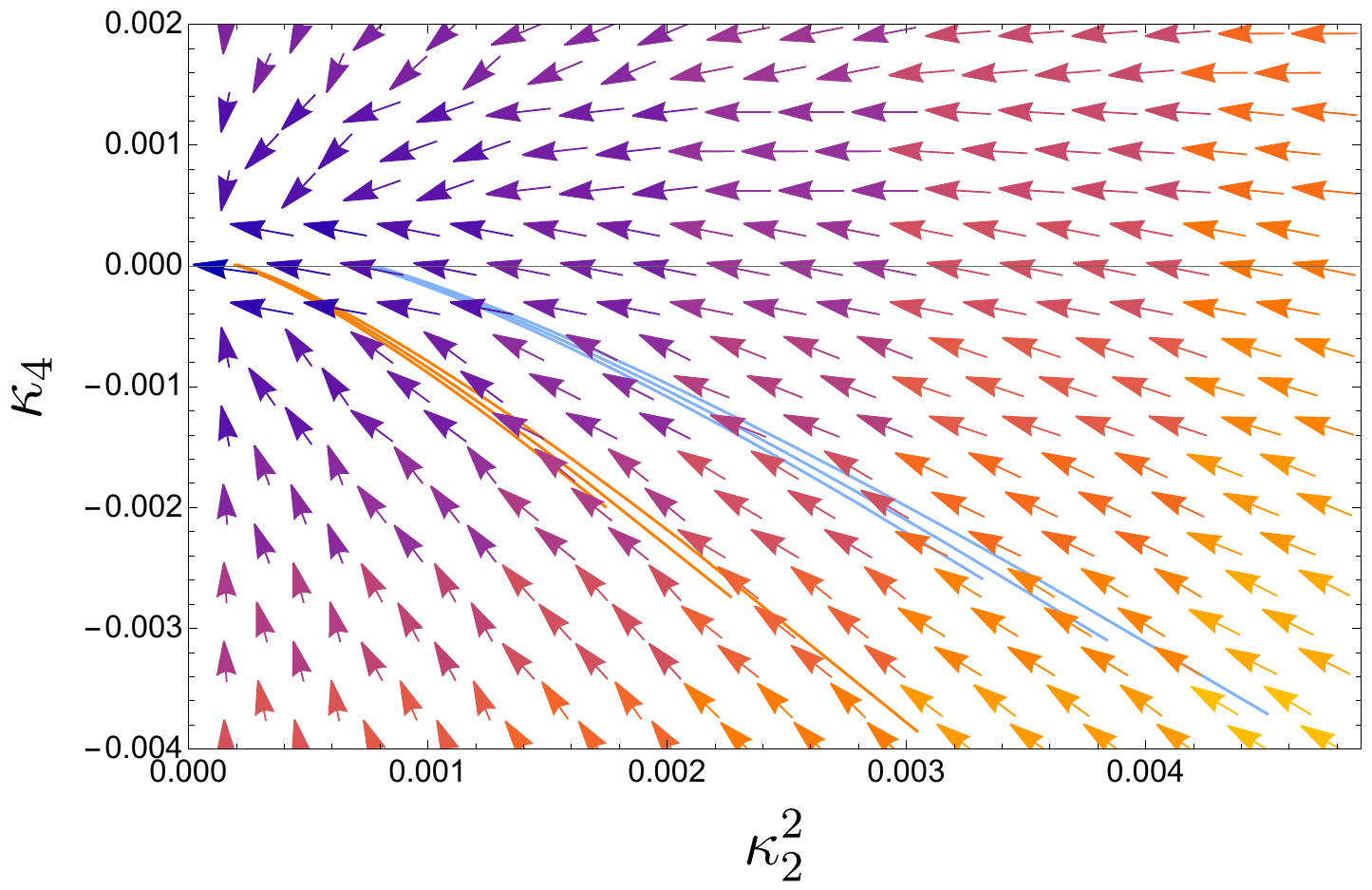}
\caption{The renormalization group flows of the cumulants $\kappa_4$ and $\kappa_2 \kappa_2$ of the $x_{\text{track}}$ distribution. The arrows indicate the direction of the flow towards the UV. This flow exhibits an interesting mixing between $\kappa_4 \to \kappa_2 \kappa_2$, which probes non-linear renormalization group equations in QCD. The specific non-perturbative parameters of QCD select out the blue and orange flows for quark and gluon jets, respectively.}\label{fig:flow_a}
\end{figure}

\subsection{Renormalization Group Flows}\label{sec:LHC_flows}

The incredible energy reach of the LHC enables measurements of the track functions over a wide range of energies, making tests of their RG flow possible. Unlike standard fragmentation functions, which exhibit a linear Dokshitzer-Gribov-Lipatov-Altarelli-Parisi (DGLAP) \cite{Gribov:1972ri,Dokshitzer:1977sg,Altarelli:1977zs} evolution, the track functions exhibit non-linear RG evolution, which results in non-trivial RG flows in the space of their moments \cite{Jaarsma:2022kdd}.  As shown in \Sec{sec:mom}, this non-linear RG evolution is directly reflected in the RG evolution of the track jet function. Therefore, one can study this evolution by measuring moments of the $x_{\text{trk}}$ distribution in high-energy jets. While the evolution of average quantities, such as $\langle x_{\text{trk}} \rangle$ are well understood, fluctuations, $\langle (x_\text{trk}-\langle x_\text{trk} \rangle)^n \rangle$ are much less understood. They have long been of interest in the study of QCD, see e.g.~ref.~\cite{Dokshitzer:1992df}. As compared to multiplicity fluctuations, which are difficult both experimentally and theoretically, fluctuations in the track energy fraction can be accurately measured, and their evolution can be computed systematically in perturbation theory. It is precisely these fluctuations that make jet substructure interesting from a QFT perspective.

Higher moments of the track energy fraction, $\langle (x_\text{trk}-\langle x_\text{trk} \rangle)^n \rangle$, provide clean experimental observables where interesting RG phenomena can occur. In particular, they exhibit mixing between products of different moments, giving rise to non-trivial RG flows. We are unaware of other examples where such phenomena can be directly measured, making measurements of higher moments of the $x_{\text{trk}}$ distribution a unique probe of Lorentzian QFT and the renormalization group in particular.

\begin{figure}
\centering
\includegraphics[width=0.75\textwidth]{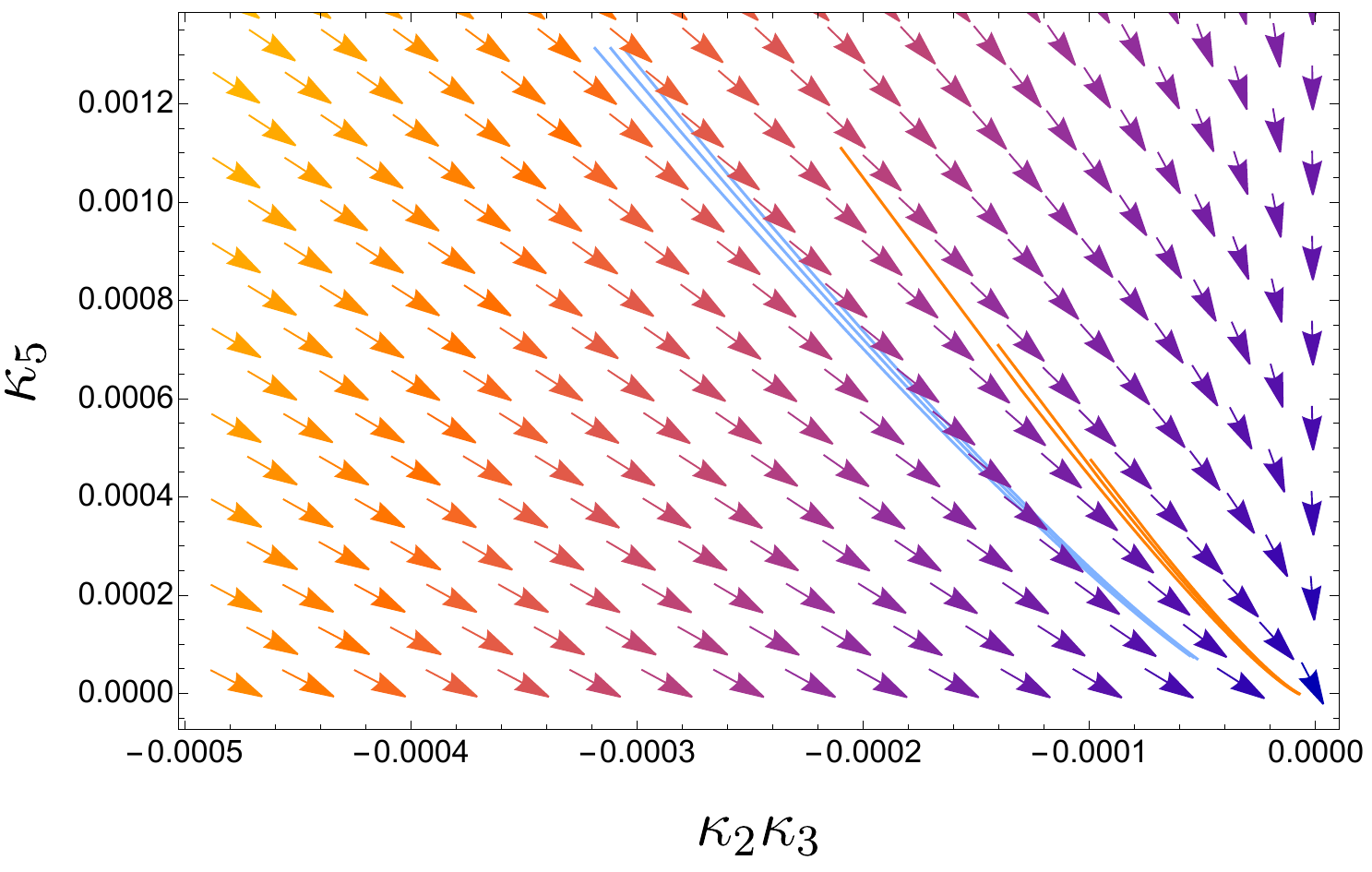}
\caption{The renormalization group flows of the cumulants $\kappa_5$ and $\kappa_3 \kappa_2$ of the $x_{\text{track}}$ distribution. The arrows indicate the direction of the flow towards the UV. This demonstrates the mixing between $\kappa_5 \to \kappa_3 \kappa_2$, probing interesting non-linear renormalization group equations in QCD. The specific non-perturbative parameters of QCD select out the blue and orange flows for quark and gluon jets, respectively.}\label{fig:flow_b}
\end{figure}

\begin{figure}
\centering
\includegraphics[width=0.75\textwidth]{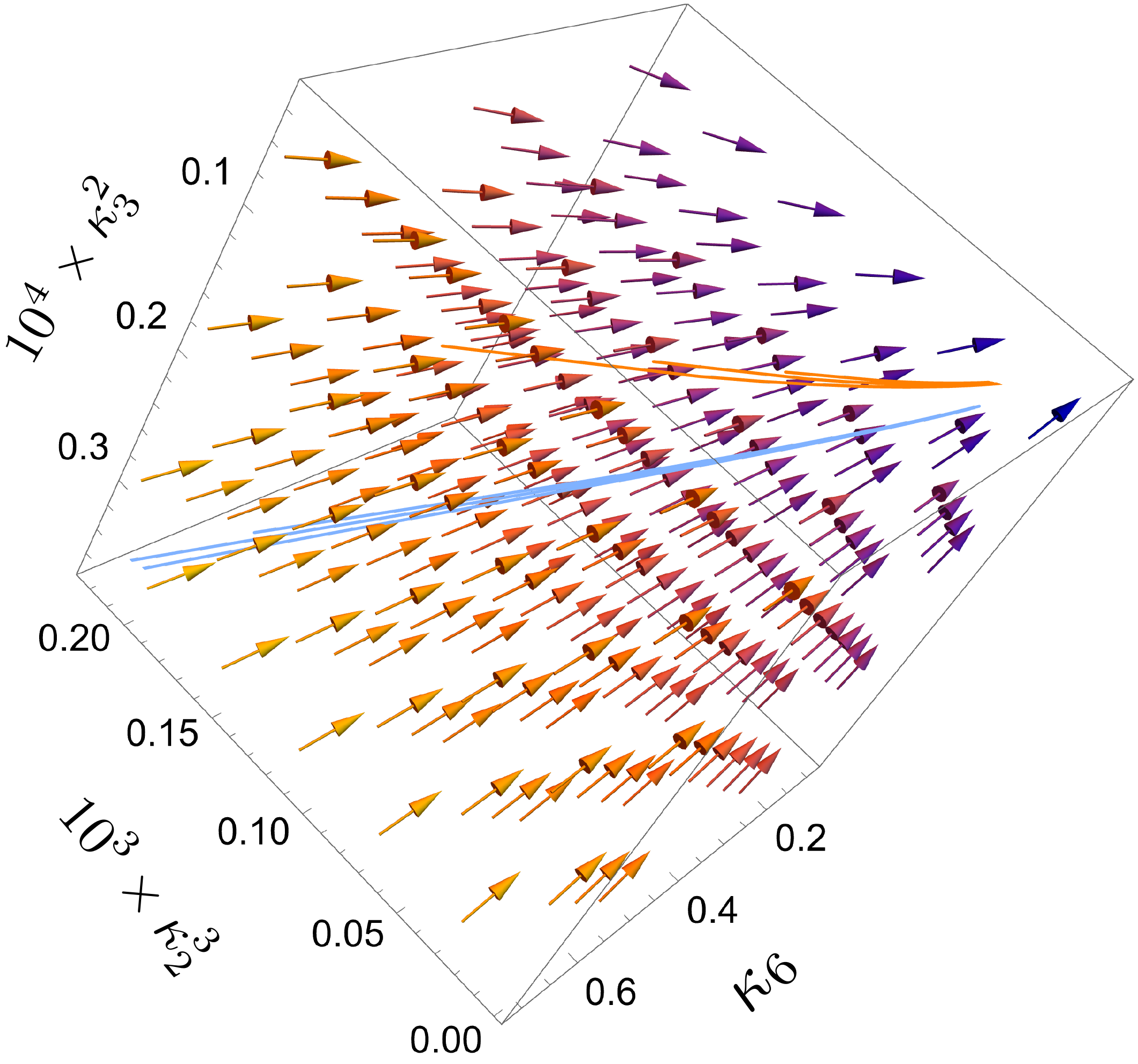}
\caption{The $3$-dimensional renormalization group flows of the cumulants $\kappa_6, \kappa_3^2, $and $\kappa_2^3$ of the $x_{\text{track}}$ distribution. This is the first moment that mixes with a product of three track functions. The specific non-perturbative parameters of QCD select out the blue and orange flows for quark and gluon jets, respectively. }\label{fig:flow_c}
\end{figure}

We present our results for the mixing of the  $x_{\text{trk}}$ moments in terms of the cumulants, $\kappa_n$ of the track function distribution. Recall that for a random variable $X$, we define the central moments as
\begin{align}
\mu_n =\text{E}[(X-\text{E}[X])^n]\,,
\end{align}
and the cumulants as the coefficients in the power series
\begin{align}
K(t)=\log \text{E}[e^{tX}]=\sum\limits_{n=1}^\infty \kappa_n \frac{t^n}{n!}\,.
\end{align}
Explicitly, for the first few central moments, we have
\begin{align}
\kappa_2&=\mu_2\,, \nn \\
\kappa_3&=\mu_3\,, \nn \\
\kappa_4&=\mu_4-3\mu_2^2\,, \nn \\
\kappa_5&=\mu_5-10 \mu_3 \mu_2\,, \nn \\
\kappa_6&=\mu_6-15\mu_4 \mu_2-10\mu_3^2+30\mu_2^3\,. 
\end{align}
The first interesting mixings arise between $\kappa_4$ and $\kappa_2 \kappa_2$, as well as between $\kappa_5$ and $\kappa_3 \kappa_2$. At $\kappa_6$, one can have for the first time a mixing into a product of three moments $\kappa_6 \to \kappa_2 \kappa_2 \kappa_2$, as well as the mixings into $\kappa_4 \kappa_2$ and $\kappa_3 \kappa_3$.

The flow of the moments of the track energy fraction in LHC jets is shown in \Fig{fig:flow_a} for $\kappa_4 \to \kappa_2 \kappa_2$ and \Fig{fig:flow_b} for $\kappa_5 \to \kappa_3 \kappa_2$.  In \Fig{fig:flow_c} we show a three dimensional mixing involving $\kappa_6$, $\kappa_3^2$ and $\kappa_2^3$.\footnote{Technically, the RG flow space is bigger than just the dimensions of the plots here. That is, the cumulants are actually given separately for quark and gluon tracks, and $\kappa_6$ also has mixing with $\kappa_4 \kappa_2$ as well. In order to represent the RG mixing in $2$- or $3$-dimensional plots shown here, we made some simple choices which cause a slight deviation between the vector fields and the actual flow of the cumulants to UV. This figure is mostly intended for illustrative purposes to highlight the interesting mixing structures.} This is the first moment where there is a triple mixing $\kappa_6\to \kappa_2^3$. In these plots we show the general structure of the RG flows using the NLO evolution equations of the track \emph{jet} functions. We also highlight the specific flows singled out by the non-perturbative parameters of QCD, shown for quark jets in blue, and gluon jets in orange evolved between $5 \text{ GeV}< p_T < 3 \text{ TeV}$. The arrows indicate the direction of the flow towards the UV. The UV fixed point is $\kappa_i=0$ for all $i >1$, corresponding to the flow  of the track function towards a delta function~\cite{Jaarsma:2022kdd}. 
These flows directly reflect the non-linear RG flows of the track functions in a physical jet substructure observable, which we find to be quite remarkable.

\subsection{HERA and the EIC}\label{sec:pheno_DIS}

\begin{figure}
\centering
\includegraphics[width=0.7\textwidth]{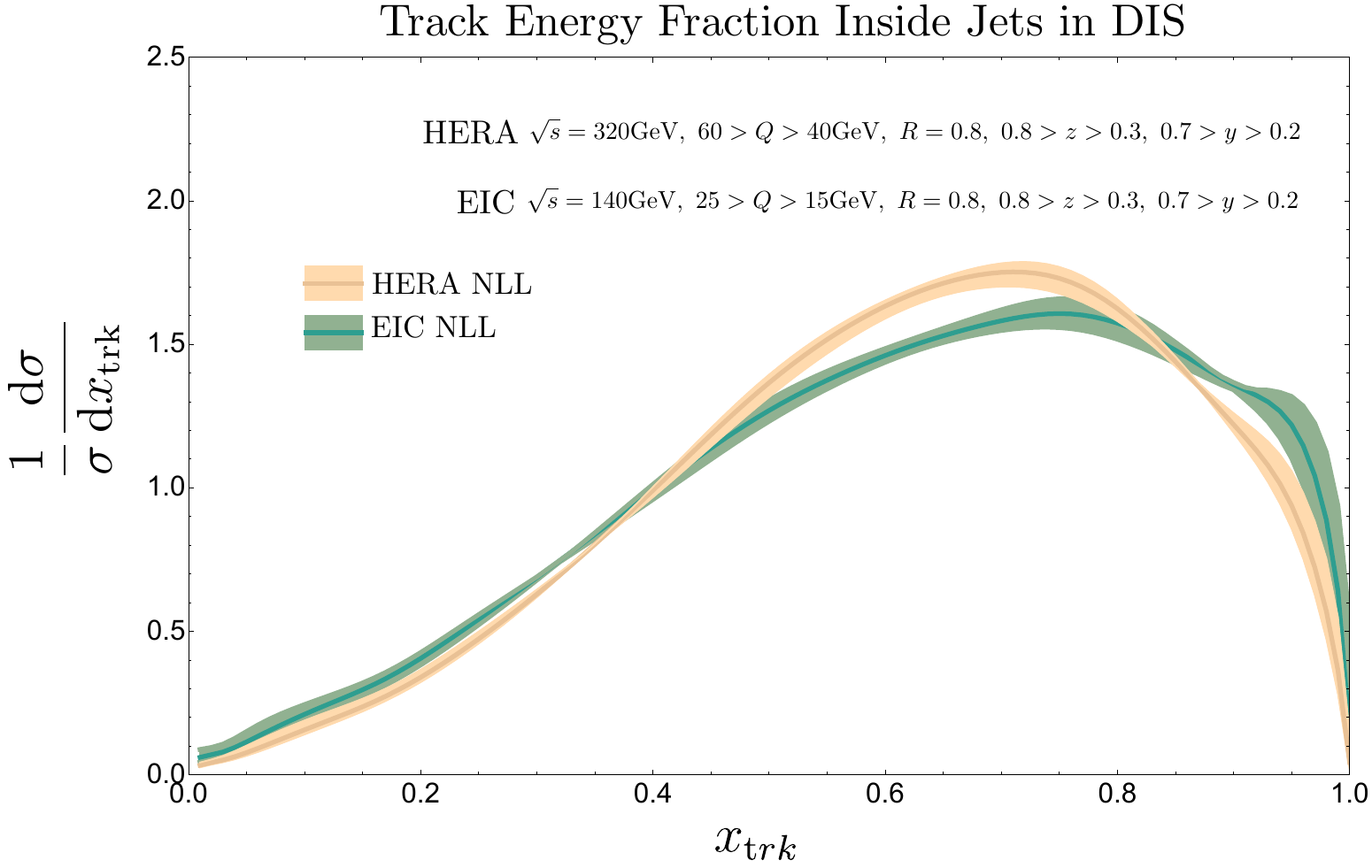}
\caption{The NLL $x_{\text{trk}}$ distribution in DIS for both HERA and EIC kinematics. In both cases, the result is dominated by quark jets. This will allow the extraction of the quark track functions from jet measurements at HERA and the EIC. }\label{fig:HERA_final}
\end{figure}

In addition to the LHC, we can also apply our formalism to extract track functions from identified jets in DIS experiments. The motivation for this is two-fold: first, there has been recent excitement in applications of re-analyzed HERA data (see e.g.~ref.~\cite{H1:2023fzk}), and it would be particularly interesting to apply this to the measurement of track functions. Second, the EIC is currently being built, which will enable extremely high-precision measurements of QCD jets in DIS processes. One potential advantage of DIS experiments, apart from their clean environment, is that it is possible to scan the flavor composition of the jets by tuning the value of Bjorken $x$. This may enable a more detailed measurement of the flavor dependence of track functions for light quarks as compared to what is possible at the LHC, which can then also be used as an input for precision studies at the LHC. This could be further enhanced by using approaches to tag the flavor of jets~\cite{Lee:2022kdn,Kang:2020fka}. However, unlike the LHC, HERA/EIC do not provide as many gluon jets.

We consider jets produced in electron-proton collisions $ep\to e+{\rm jet}+X$. The photon virtuality $Q^2$ and Bjorken $x$ are determined by measuring the scattered electron. Here we limit ourselves to the high-$Q^2$ DIS region. At LO, the only contribution is the electron-quark scattering channel. In order to capture the LO process, we consider jets reconstructed in the Breit frame using a spherically invariant anti-$k_T$ algorithm similar to the one used in $e^+e^-$ collisions~\cite{Arratia:2020ssx}. Instead of requiring large transverse momentum jets, we measure the jet energy $z$. The cross section including the jet substructure measurement $x_{\rm trk}$, can be written as
\begin{align}
\frac{\mathrm{d} \sigma}{\mathrm{d} x\,\mathrm{d} Q^2\,\mathrm{d} z\, \mathrm{d} x_{\rm trk}}= \mathcal{H}^{\rm DIS}_{i}(x,z,Q, \mu)\otimes_z  \mathcal{G}_{i\to \rm{trk}}(z,x_{\rm trk},Q R,\mu)\,.
\end{align}
As in \eqref{eq:fact}, a sum over repeated indices is implied and $\otimes_z$ denotes a convolution integral over the jet energy fraction $z$. Note that the hard scale of the process is set by the photon virtuality $Q^2$ instead of the jet transverse momentum as in $pp$ collisions discussed above. Using the relation $Q^2=xys$, we also include a cut on the inelasticity $y$. In order to evaluate the cross section, we make use of a double Mellin transform for the variables $x,z$. The NLO hard functions $\mathcal{H}^{\rm DIS}_{i}$ in Mellin space can be found in ref.~\cite{Anderle:2012rq}. In \Fig{fig:HERA_final} we show the $x_{\text{trk}}$ distribution at HERA and EIC. Again, we observe small uncertainties at NLL despite the relatively low energies of HERA/EIC. Finally, we note that, analogous to $e^+e^-$ collisions, track functions could also be accessed in $ep$ collisions without the reconstruction of jets, since $Q$ provides a jet-independent hard scale.

\section{Conclusions}\label{sec:conc}

The use of tracking information will play an increasingly important role in future developments in jet substructure. Due to recent improvements in our understanding of the RG evolution of track functions and their application to physical observables, we believe that it is timely to extract them from data. This is made complicated at hadron colliders due to the necessity to first identify hard jets using jet algorithms, introducing jet-algorithm dependence into any observable used to extract the track functions.

In this paper we have derived the factorization theorem required to extract track functions from measurements of the track energy fraction in high energy jets at hadron colliders. This involved the introduction of a new ``track jet function", which can be matched onto the universal track functions, with the perturbatively-calculable matching coefficient incorporating all aspects of the jet algorithm. We studied the basic properties of this observable, including its renormalization group evolution, computed its matching onto track functions at one loop, and determined the logarithmically enhanced two-loop terms.

We applied these results to derive phenomenological predictions for measurements of the track energy fraction in jets at both the LHC, and at HERA/EIC. These illustrated the large impact of the matching coefficients appearing in the expansion of the track jet functions onto track functions for properly extracting the universal track functions from jet-based measurements. Our results enable the extraction of universal track functions from hadron collider measurements, opening the door to their application in a wide variety of jet substructure measurements. It would be of particular interest to extract the track functions from two different colliders with widely different energies, for example the LHC and HERA, to test their universality and renormalization group properties. 

The precision measurement of the track function will be useful both as a foundational non-perturbative quantity that cannot currently be computed from first principles in QCD, but also as a powerful new test of QCD through its perturbatively calculable renormalization group evolution. From a theoretical perspective, a particularly interesting aspect of the track functions is that they exhibit a non-linear evolution equation, characteristic of Lorentzian physics. In moment space, this translates to a mixing between moments of the track functions, and products of moments, leading to interesting Lorentzian renormalization group flows. These are directly imprinted into the physical $x_\text{trk}$ observable, allowing these flows to be cleanly measured at the LHC. We find this to be quite remarkable, and we are unaware of other examples where such physics can be accessed in a similar manner. Their measurement will provide significant insight into non-linear renormalization group equations in QCD.

The precision measurement of track functions will open a new era in jet substructure, enabling data-theory comparisons for a much broader class of jet substructure observables, and allowing us to probe increasingly subtle features in the substructure of jets, revealing clues about the microscopic dynamics of QCD and the Standard Model.

\acknowledgments
We thank Matt LeBlanc, Ben Nachman, Jennifer Roloff, and in particular Jingjing Pan, for extensive discussions on the extraction of track functions at the LHC, which motivated this work, as well as for guidance with the experimental literature. We thank Duff Neill and Xiaoyuan Zhang for useful discussions. K.L.~was supported by the U.S.~DOE under contract number DE-SC0011090. I.M.~is supported by start-up funds from Yale University. F.R. is supported by the DOE with contract No.~DE-AC05-06OR23177, under which Jefferson Science Associates, LLC operates Jefferson Lab. W.W.~is supported by the D-ITP consortium, a program of NWO that is funded by the Dutch Ministry of Education, Culture and Science (OCW).

\bibliography{fragmenting_track.bib}
\bibliographystyle{JHEP}

\end{document}